\documentclass{article} \newcommand{\figsize}{2.8}
\usepackage{spconf}
\ninept

\usepackage{graphicx,psfrag}
\usepackage[cmex10]{amsmath}
\usepackage{amsfonts,amssymb,latexsym,hyperref}
\usepackage{cite}
\usepackage[usenames,dvipsnames]{color}

\usepackage{etoolbox}
\preto\subequations{\ifhmode\unskip\fi}

 \newcommand{\putFrag}[4]{\begin{figure}[t]
                            \centering
                            #4
                            \vspace{3mm}
                            \includegraphics[width=#3in]{figures/#1.eps}
                            \vspace{-1mm}
                            \caption{#2}
                            \vspace{-2mm}
                            \label{fig:#1}
                          \end{figure} }
 
 \newcommand{\capFrag}[2]{}
 \newcommand{\capTable}[2]{}

 \renewcommand{\tilde}{\widetilde}
 \renewcommand{\hat}{\widehat}

 \newcommand{\defn}{\triangleq}

 \newcommand{\hvec}[1]{\ensuremath{\Hat{\boldsymbol{#1}}}}
 
 \renewcommand{\vec}[1]{\ensuremath{\boldsymbol{#1}}}

 \newcommand{\norm}[1]{\ensuremath{\| #1 \|}}
 \newcommand{\mc}[1]{\ensuremath{\mathcal{#1}}}

 \newcommand{\Real}{{\mathbb{R}}}
 \newcommand{\Complex}{{\mathbb{C}}}

 \newcommand{\of}[1]{^{(#1)}}

 \newcommand{\herm}{^{\text{\textsf{H}}}}

 \DeclareMathOperator{\sgn}{sgn}
 
 \DeclareMathOperator{\E}{E}

 \DeclareMathOperator{\tr}{tr}

 \renewcommand{\eqref}[1]{(\ref{eq:#1})}
 
 \newcommand{\Figref}[1]{Figure~\ref{fig:#1}}
 \newcommand{\figref}[1]{Fig.~\ref{fig:#1}}
 
 \newcommand{\secref}[1]{Section~\ref{sec:#1}}

 \newcounter{comment}[section]
 
 \newcounter{texthead}[section]

\begin{document}
\topmargin=0mm
\setlength{\arraycolsep}{0.8mm}

\title{Onsager-Corrected Deep Learning for Sparse Linear Inverse Problems}

\name{
Mark Borgerding and Philip Schniter
}
\address{
Dept. of ECE, The Ohio State University, Columbus, OH 43202\\
Email: borgerding.7@osu.edu, schniter.1@osu.edu
}

\maketitle

\begin{abstract}
Deep learning has gained great popularity due to its widespread success on many inference problems.
We consider the application of deep learning to the sparse linear inverse problem encountered in compressive sensing, where one seeks to recover a sparse signal from a small number of noisy linear measurements.
In this paper, we propose a novel neural-network architecture that decouples prediction errors across layers in the same way that the approximate message passing (AMP) algorithm decouples them across iterations: through Onsager correction.
Numerical experiments suggest that our ``learned AMP'' network significantly improves upon Gregor and LeCun's ``learned ISTA'' network in both accuracy and complexity.\footnote{This work was supported by the NSF I/UCRC grant IIP-1539960.}
\end{abstract}



%

\section{Introduction}

We consider the problem of recovering a signal $\vec{s}\in\Complex^N$ from a noisy linear measurement 
\begin{align}
\vec{y} = \vec{\Phi s} + \vec{n} \in\Complex^M 
\label{eq:ys} ,
\end{align}
where $\vec{\Phi}\in\Complex^{M\times N}$ represents a linear measurement operator and $\vec{n}\in\Complex^M$ a noise vector.
In many cases of interest, $M\ll N$.
We will assume that the signal vector $\vec{s}$ has a sparse representation in a known orthonormal basis $\vec{\Psi}\in\Complex^{N\times N}$, i.e., that $\vec{s}=\vec{\Psi x}$ for some sparse vector $\vec{x}\in\Complex^N$.
Thus we define $\vec{A}\defn\vec{\Phi \Psi}\in\Complex^{M\times N}$, write \eqref{ys} as
\begin{align}
\vec{y} = \vec{Ax} + \vec{n} 
\label{eq:y} ,
\end{align}
and seek to recover an sparse $\vec{x}$ from $\vec{y}$.
In the sequel, we will refer to this problem as the ``sparse linear inverse'' problem.
Note that the resulting estimate $\hvec{x}$ of $\vec{x}$ can be easily converted into an estimate $\hvec{s}$ of $\vec{s}$ via $\hvec{s}=\vec{\Psi}\hvec{x}$.

The sparse linear inverse problem has received enormous attention over the last few years, in part because it is central to compressive sensing \cite{Eldar:Book:12} and sparse coding \cite{Olshausen:VR:97}.
Many methods have been developed to solve this problem. 
Most of the existing methods involve a reconstruction \emph{algorithm} that inputs a pair $(\vec{y},\vec{A})$ and produces an sparse estimate $\hvec{x}$.
A myriad of such algorithms have been proposed,
including both sequential (e.g., greedy) and iterative varieties.
Some relevant algorithms will be reviewed in Section \secref{algorithms}.

Recently, a different approach to solving this problem has emerged from the field of ``deep learning,'' whereby a neural network with many layers is trained using a set of $D$ examples\footnote{%
Since orthonormal $\vec{\Psi}$ implies $\vec{x}=\vec{\Psi}\herm\vec{s}$, training examples of the form $\{(\vec{y}\of{d},\vec{s}\of{d})\}$ can be converted to $\{(\vec{y}\of{d},\vec{x}\of{d})\}_{d=1}^D$ via $\vec{x}\of{d}=\vec{\Psi}\herm\vec{s}\of{d}$.} 
$\{(\vec{y}\of{d},\vec{x}\of{d})\}_{d=1}^D$.
Once trained, the network can be used to predict the sparse $\vec{x}$ that corresponds to a given input $\vec{y}$.
Note that knowledge of the operator $\vec{A}$ is not necessary here.
Previous work (e.g., 
\cite{Gregor:ICML:10, 
Sprechmann:ICML:12, 
Kamilov:SPL:16, 
Wang:AAAI:16}
)
has shown that the deep-learning approach to solving sparse linear inverse problems has the potential to offer significant improvements, in both accuracy and complexity, over the traditional algorithmic approach.

\emph{Relation to prior work:}
In this paper, we show how recent advances in iterative reconstruction algorithms suggest modifications to traditional neural-network architectures that yield improved accuracy and complexity when solving sparse linear inverse problems.
In particular, we show how ``Onsager correction,'' which lies at the heart of the approximate message passing (AMP) algorithm \cite{Donoho:PNAS:09}, can be employed to construct deep networks with increased accuracy and computational efficiency (i.e., fewer layers needed to produce an accurate estimate).
To our knowledge, the application of Onsager correction to deep neural networks is novel.

\section{Iterative Algorithms and Deep Learning} \label{sec:background}

\subsection{Iterative Algorithms} \label{sec:algorithms}

One of the best known algorithmic approaches to solving the sparse linear inverse problem is through solving the convex optimization problem \cite{Tibshirani:JRSSb:96,Chen:JSC:98} 
\begin{align}
\hvec{x} 
&= \arg\min_{\vec{x}} \tfrac{1}{2} \|\vec{y}-\vec{Ax}\|_2^2 + \lambda \|\vec{x}\|_1 
\label{eq:lasso} ,
\end{align}
where $\lambda >0$ is a tunable parameter that controls the tradeoff between sparsity and measurement fidelity in $\hvec{x}$.
The convexity of \eqref{lasso} leads to provably convergent algorithms and bounds on the performance of the estimate $\hvec{x}$ (see, e.g., \cite{Candes:CPAM:06}).

\subsubsection{ISTA} \label{sec:ista}

One of the simplest approaches to solving \eqref{lasso} is the iterative soft-thresholding algorithm (ISTA) \cite{Chambolle:TIP:98}, which consists of iterating the steps (for $t=0,1,2,\dots$ and $\hvec{x}_0=\vec{0}$)
\begin{subequations} \label{eq:ista}
\begin{align}
\vec{v}_t 
&= \vec{y} - \vec{A}\hvec{x}_t \\
\hvec{x}_{t+1} 
&= \eta \big( \hvec{x}_t + \beta\vec{A}\herm \vec{v}_t; \lambda \big) ,
\end{align}
\end{subequations}
where $\beta\in(0,1/\|\vec{A}\|^2_2]$ is a stepsize,
$\vec{v}_t$ is the iteration-$t$ residual measurement error, and
$\eta(\cdot;\lambda):\Complex^N\rightarrow\Complex^N$ is the ``soft thresholding'' denoiser that operates componentwise as:
\begin{align}
[\eta(\vec{r};\lambda)]_j 
&= \sgn(r_j) \max\{|r_j|-\lambda,0\} 
\label{eq:soft_thresh} .
\end{align}

\subsubsection{FISTA} \label{sec:fista}

Although ISTA is guaranteed to converge under $\beta\in(0,1/\|\vec{A}\|^2_2)$ \cite{Daubechies:CPAM:04}, it converges somewhat slowly and so many modifications have been proposed to speed it up.
Among the most famous is ``fast ISTA'' (FISTA) \cite{Beck:JIS:09},
\begin{subequations} \label{eq:fista}
\begin{align}
\vec{v}_t 
&= \vec{y} - \vec{A}\hvec{x}_t \\
\hvec{x}_{t+1} 
&= \eta \big( 
   \hvec{x}_t 
   + \beta\vec{A}\herm  \vec{v}_{t} 
   + \tfrac {t-2} {t+1} \left( \hvec{x}_t - \hvec{x}_{t-1} \right) 
   ; \lambda \big) ,
\end{align}
\end{subequations}
which converges in roughly an order-of-magnitude fewer iterations than ISTA (see \figref{ista_fista_amp_phil}).

\subsubsection{AMP} \label{sec:amp}

Recently, the approximate message passing (AMP) algorithm \cite{Donoho:PNAS:09,Montanari:Chap:12} was applied to \eqref{lasso}, giving
\begin{subequations} \label{eq:amp}
\begin{eqnarray}
\vec{v}_t 
&=& \vec{y} - \vec{A}\hvec{x}_t + b_t \vec{v}_{t-1} 
\label{eq:amp1} \\
\hvec{x}_{t+1} 
&=& \eta \big( \hvec{x}_t + \vec{A}\herm\vec{v}_t; \lambda_t \big), 
\label{eq:amp2} 
\end{eqnarray}
\end{subequations}
for $\hvec{x}_0=\vec{0}$, $\vec{v}_{-1}=\vec{0}$, $t=0,1,2,\dots$, and
\begin{align}
b_t 
&= \tfrac{1}{M}\|\hvec{x}_t\|_0
\label{eq:bt} \\
\lambda_t 
&= \tfrac{\alpha}{\sqrt{M}} \norm{\vec{v}_t}_2
\label{eq:lambda} .
\end{align}
Here, $\alpha$ is a tuning parameter that has a one-to-one correspondence with $\lambda$ in \eqref{lasso} \cite{Montanari:Chap:12}.
Comparing AMP to ISTA, we see two major differences: 
i) AMP's residual $\vec{v}_t$ in \eqref{amp1} includes the ``Onsager correction'' term $b_t\vec{v}_{t-1}$, and
ii) AMP's denoising threshold $\lambda_t$ in \eqref{amp2} takes the prescribed, $t$-dependent value \eqref{lambda}.
We now describe the rationale behind these differences.

When $\vec{A}$ is a typical realization of a large i.i.d.\ (sub)Gaussian random matrix with entries of variance $M^{-1}$, the Onsager correction \emph{decouples} the AMP iterations in the sense that the input to the denoiser,
$
\vec{r}_t 
\defn \hvec{x}_t + \vec{A}\herm\vec{v}_t
$,
can be modeled as\footnote{%
The AMP model \eqref{lsl} is provably accurate in the large-system limit (i.e., $M,N\rightarrow\infty$ with $M/N$ converging to a fixed positive constant) \cite{Bayati:TIT:11}.
} 
\begin{align} 
\vec{r}_t
&= \vec{x} + \mc{N}(\vec{0},\sigma_t^2\vec{I}_N) 
~~\text{with}~~
\sigma_t^2  
= \tfrac{1}{M}\|\vec{v}_t\|_2^2 .
\label{eq:lsl}
\end{align}
In other words, the Onsager correction ensures that the denoiser input is an additive white Gaussian noise (AWGN) corrupted version of the true signal $\vec{x}$ with known AWGN variance $\sigma_t^2$.
(See \figref{qqplots}.)
The resulting problem, known as ``denoising,'' is well understood.
For example, for an independent known prior $p(\vec{x})=\prod_{j=1}^N p_j(x_j)$, the mean-squared error (MSE)-optimal denoiser\footnote{AMP with MSE-optimal denoising was first described in \cite{Donoho:ITW:10a}.} is simply the posterior mean estimator
(i.e., $\hat{x}_{t+1,j}=\E\{x_j|r_{t,j};\sigma_t\}$),
which can be computed in closed form for many $p_j(\cdot)$.
In the more realistic case that $p_j(\cdot)$ are unknown, we may be more interested in the \emph{minimax} denoiser, i.e., minimizer of the maximum MSE over an assumed family of priors.
Remarkably, for sparse priors, i.e., $p_j(x_j) = (1-\gamma)\delta(x_j)+\gamma \tilde{p}_j(x_j)$ with $\gamma\in(0,1)$ and arbitrary unknown $\tilde{p}_j(\cdot)$,
soft-thresholding \eqref{soft_thresh} with a threshold proportional to the AWGN standard deviation (i.e., $\lambda_t=\alpha\sigma_t$ as in \eqref{lambda}) is nearly minimax optimal \cite{Montanari:Chap:12}.
Thus, we can interpret the AMP algorithm \eqref{amp} as a nearly minimax approach to the sparse linear inverse problem.

\subsubsection{Comparison of ISTA, FISTA, and AMP} \label{sec:compare}

We now compare the average per-iteration behavior of ISTA, FISTA, and AMP 
for an $\vec{A}$ drawn i.i.d.\ $\mc{N}(0,M^{-1})$.
In our experiment,
all quantities were real-valued, 
the problem dimensions were $N=500$ and $M=250$,
the elements of $\vec{x}$ were drawn i.i.d.\ $\mc{N}(0,1)$ with probability $\gamma=0.1$ and were otherwise set to zero, and 
the noise $\vec{n}$ was drawn i.i.d.\ $\mc{N}(0,v)$ with $v$ set to yield a signal-to-noise ratio (SNR) of $40$~dB.
Recall that ISTA, FISTA, and AMP all estimate $\vec{x}$ by iteratively minimizing \eqref{lasso} for a chosen value of $\lambda$ (selected via $\alpha$ in the case of AMP). 
We chose the minimax optimal value of $\alpha$ for AMP (which is $1.1402$ since $\gamma=0.1$ \cite{Montanari:Chap:12}) and used the corresponding $\lambda$ for ISTA and FISTA.
\Figref{ista_fista_amp_phil} shows the average normalized MSE (NMSE) versus iteration $t$, where NMSE $\defn \|\hvec{x}_t-\vec{x}\|_2^2/\|\vec{x}\|_2^2$ and $1000$ realizations of $(\vec{x},\vec{n})$ were averaged. 
We see that AMP requires roughly an order-of-magnitude fewer iterations than FISTA, which requires roughly an order-of-magnitude fewer iterations than ISTA.

\putFrag{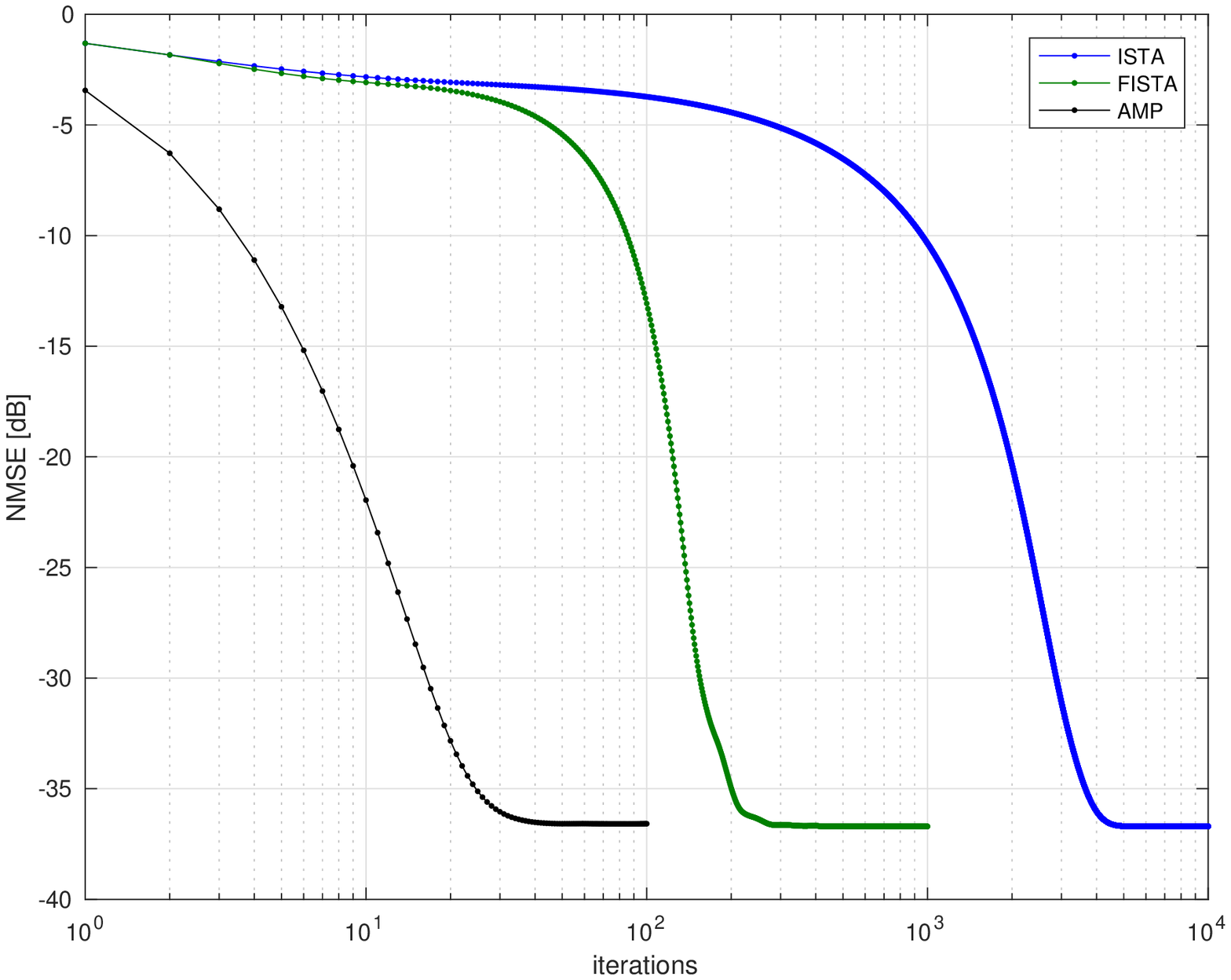}
   {Average NMSE versus iteration number for AMP, FISTA, ISTA (from left to right).  
   }
   {\figsize}
   {\psfrag{ISTA kappa=15}[l][l][0.55]{\hspace{-1pt}\sf ISTA $\kappa=15$} 
    \psfrag{FISTA kappa=15}[l][l][0.55]{\hspace{-1pt}\sf FISTA $\kappa=15$}
    \psfrag{AMP kappa=15}[l][l][0.55]{\hspace{-1pt}\sf AMP $\kappa=15$}
    \psfrag{Genie kappa=15}[l][l][0.55]{\hspace{-1pt}\sf Oracle $\kappa=15$}
    \psfrag{ISTA Gaussian}[l][l][0.55]{\hspace{-1pt}\sf ISTA i.i.d.\ $\mc{N}$} 
    \psfrag{FISTA Gaussian}[l][l][0.55]{\hspace{-1pt}\sf FISTA i.i.d.\ $\mc{N}$}
    \psfrag{AMP Gaussian}[l][l][0.55]{\hspace{-1pt}\sf AMP i.i.d.\ $\mc{N}$}
    \psfrag{Genie Gaussian}[l][l][0.55]{\hspace{-1pt}\sf Oracle i.i.d.\ $\mc{N}$}
    \psfrag{Recovery NMSE (dB)}[B][][0.7]{\sf average NMSE [dB]}
    \psfrag{NMSE [dB]}[][][0.7]{\sf average NMSE [dB]}
    \psfrag{iterations}[][][0.7]{\sf iterations}
   }

\subsection{Deep Learning} \label{sec:learning}

In deep learning \cite{Goodfellow:Book:16}, training data $\{(\vec{y}\of{d},\vec{x}\of{d})\}_{d=1}^D$ composed of (feature,label) pairs are used to train the parameters of a deep neural network with the goal of accurately predicting the label $\vec{x}$ of a test feature $\vec{y}$.
The deep network accepts $\vec{y}$ and subjects it to many layers of processing, where each layer consists of a linear transformation followed by a non-linearity.

Typically, the label space is discrete (e.g., $\vec{y}$ is an image and $\vec{x}$ is its class in \{cat, dog,\dots, tree\}).
In our sparse linear inverse problem, however, the ``labels'' $\vec{x}$ are continuous and high-dimensional (e.g., $\Complex^N$ or $\Real^N$). 
Remarkably, Gregor and LeCun demonstrated in \cite{Gregor:ICML:10} that a well-constructed deep network can accurately predict even labels such as ours.

The neural network architecture proposed in \cite{Gregor:ICML:10} is closely related to the ISTA algorithm discussed in \secref{ista}.
To understand the relation, we rewrite the ISTA iteration \eqref{ista} as
\begin{align}
\hvec{x}_{t+1} 
&= \eta \big( \vec{S}\hvec{x}_t + \vec{B}\vec{y}; \lambda \big) 
\text{~~with~} 
\begin{cases}
\vec{B}\defn \beta\vec{A}\herm \\
\vec{S}\defn \vec{I}_N-\vec{B}\vec{A} 
\end{cases}
\label{eq:ista2}
\end{align}
and ``unfold'' iterations $t=1...T$, resulting in the $T$-layer feed-forward neural network shown in \figref{lista_3layer}. 
Whereas ISTA uses the values of $\vec{S}$ and $\vec{B}$ prescribed in \eqref{ista2} and a common value of $\lambda$ at all layers, Gregor and LeCun \cite{Gregor:ICML:10} proposed to use layer-dependent thresholds $\vec{\lambda}\defn[\lambda_1,\lambda_2,\dots,\lambda_T]$ and ``learn'' both the thresholds $\vec{\lambda}$ and the matrices $\vec{B},\vec{S}$ from the training data $\{(\vec{y}\of{d},\vec{x}\of{d})\}_{d=1}^D$ by minimizing the quadratic loss 
\begin{align}
\mc{L}_T(\vec{\Theta}) = \frac{1}{D}\sum_{d=1}^D \big\|\hvec{x}_T(\vec{y}\of{d};\vec{\Theta})-\vec{x}\of{d}\big\|_2^2 .
\label{eq:loss}
\end{align}
Here, $\vec{\Theta}=[\vec{B},\vec{S},\vec{\lambda}]$ denotes the set of learnable parameters and $\hvec{x}_T(\vec{y}\of{d};\vec{\Theta})$ the output of the $T$-layer network with input $\vec{y}\of{d}$ and parameters $\vec{\Theta}$.
The resulting approach was coined ``learned ISTA'' (LISTA).

\putFrag{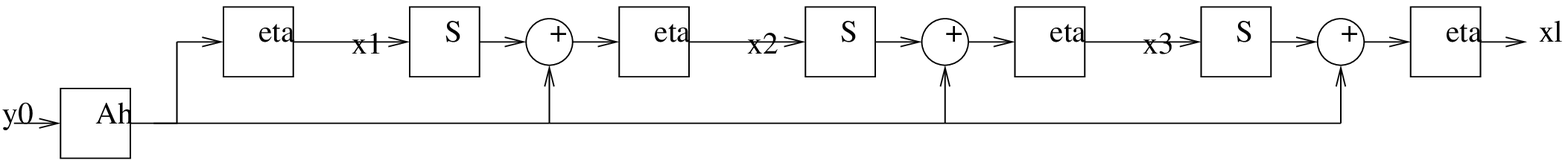}
        {The feed-forward neural network constructed by unfolding $T\!=\!4$ iterations of ISTA.} 
        {3.3}
        {\newcommand{\sz}{0.95}
         \newcommand{\szz}{0.8}
         \psfrag{+}[c][Bl][\szz]{$+$}
         \psfrag{-}[r][Bl][\szz]{$-$}
         \psfrag{x}[c][Bl][\sz]{$\times$}
         \psfrag{y0}[r][Bl][\sz]{$\vec{y}$}
         \psfrag{0}[r][Bl][\sz]{$\vec{0}$}
         \psfrag{r}[t][Bl][\sz]{$\vec{r}_t$}
         \psfrag{Ah}[c][Bl][\sz]{$\vec{B}$}
         \psfrag{S}[c][Bl][\sz]{$\vec{S}$}
         \psfrag{x1}[t][Bl][\szz]{$\hvec{x}_1$}
         \psfrag{x2}[t][Bl][\szz]{$\hvec{x}_2$}
         \psfrag{x3}[t][Bl][\szz]{$\hvec{x}_3$}
         \psfrag{xl}[l][Bl][\sz]{$\hvec{x}_4$}
         \psfrag{eta}[c][Bl][\sz]{$\eta$}
        }

Relative to existing algorithms for the sparse linear inverse problem with optimally tuned regularization parameters (e.g., $\lambda$ or $\alpha$), LISTA generates estimates of comparable MSE with significantly fewer matrix-vector multiplications.
As an example, for the 
problem described in \secref{compare}, 
LISTA took only $16$ layers to reach an NMSE of $-35$~dB, whereas AMP took $25$ iterations.
(More details will be given in \secref{numerical}.)

Other authors have also applied ideas from deep learning to the sparse linear inverse problem.
For example, \cite{Sprechmann:ICML:12} extended the approach from \cite{Gregor:ICML:10} to handle structured sparsity and dictionary learning (when the training data are $\{\vec{y}\of{d}\}_{d=1}^D$ and $\vec{A}$ is unknown).
More recently, \cite{Wang:AAAI:16} extended \cite{Gregor:ICML:10} from the $\ell_2\!+\!\ell_1$ objective of \eqref{lasso} to the $\ell_2\!+\!\ell_0$ objective,
and \cite{Kamilov:SPL:16} proposed to learn the MSE-optimal scalar denoising function $\eta$ by learning the parameters of a B-spline.
The idea of ``unfolding'' an iterative algorithm and learning its parameters via training data has also been used to recover non-sparse signals.
For example, it has been applied to 
speech enhancement \cite{Hershey:Tech:14}, 
image deblurring \cite{Schmidt:CVPR:14},
image super resolution \cite{Dong:TPAMI:16},
compressive imaging \cite{Mousavi:ALL:15,Kulkarni:CVPR:16}, 
and 
video compressive sensing \cite{Iliadis:16}.

\section{Learned AMP} \label{sec:lamp}

Recall that LISTA involves learning the matrix $\vec{S}\defn \vec{I}_N-\vec{BA}$, where $\vec{B}\in\Complex^{N\times M}$ and $\vec{A}\in\Complex^{M\times N}$.
As noted in \cite{Gregor:ICML:10}, when $M\!<\!N/2$, it is advantageous to leverage the $\vec{I}_N-\vec{BA}$ structure of $\vec{S}$, leading to network layers of the form shown in \figref{lista_1layer}, with first-layer inputs $\hvec{x}_0=\vec{0}$ and $\vec{v}_0=\vec{y}$.
Although not considered in \cite{Gregor:ICML:10}, the network in \figref{lista_1layer} allows both $\vec{A}$ and $\vec{B}$ to vary with the layer $t$, 
allowing for modest improvement (as will be demonstrated in \secref{numerical}).

\subsection{The LAMP Network}

We propose to construct a neural network from unfolded AMP \eqref{amp} with tunable parameters $\{\vec{A}_t,\vec{B}_t,\alpha_t\}_{t=0}^{T-1}$ learned from training data.
The hope is that a ``learned AMP" (LAMP) will require fewer layers than LISTA, just as AMP typically requires many fewer iterations than ISTA to converge. 

\Figref{lamp_1layer} shows one layer of the LAMP network.  
Comparing \figref{lamp_1layer} to \figref{lista_1layer}, we see two main differences:
\begin{enumerate}
\item
LAMP includes a feed-forward path from $\vec{v}_t$ to $\vec{v}_{t-1}$ that is not present in LISTA.
This path implements an ``Onsager correction'' whose goal is to \emph{decouple} the layers of the network, just as it decoupled the iterations of the AMP algorithm (recall \secref{amp}).
\item
LAMP's denoiser threshold $\lambda_t=\alpha_t\|\vec{v}_t\|_2/\sqrt{M}$ varies with the realization $\vec{v}_t$, whereas LISTA's is constant.
\end{enumerate}

Note that LAMP is built on a \emph{generalization} of the AMP algorithm \eqref{amp} wherein the matrices $(\vec{A},\vec{A}\herm)$ manifest as $(\vec{A}_t,\vec{B}_t)$ at iteration $t$.
An important question is whether this generalization preserves the independent-Gaussian nature \eqref{lsl} of the denoiser input error---the key feature of AMP.
It can be shown that the desired behavior does occur when i) $\vec{A}_t=\beta_t\vec{A}$ with scalar $\beta_t$ and ii) $\vec{B}_t=\vec{A}\herm\vec{C}_t$ with appropriately scaled $\vec{C}_t$. 
Thus, for LAMP, we impose\footnote{Here ``$\vec{A}$'' refers to the true measurement matrix from \eqref{y}.  If $\vec{A}$ is unknown, it can be learned from the training data as described in \secref{learning}.  
} 
$\vec{A}_t=\beta_t\vec{A}$ and learn only $\beta_t$, and we initialize $\vec{B}_t$ appropriately before learning.

From \figref{lamp_1layer}, we see that the $\beta_t$ scaling within $\vec{A}_t$ can be moved to the denoiser $\eta(\cdot;\cdot)$ under a suitable re-definition of $\hvec{x}_t$, and we take this approach in LAMP.
Thus, the $t$th layer of LAMP can be summarized as
\begin{subequations} \label{eq:lamp}
\begin{eqnarray}
\hvec{x}_{t+1} 
&=& \beta_t \eta \big( \hvec{x}_t + \vec{B}_t\vec{v}_t; \tfrac{\alpha_t}{\sqrt{M}} \|\vec{v}_t\|_2 \big) 
\label{eq:lamp2} \\
\vec{v}_{t+1} 
&=& \vec{y} - \vec{A}\hvec{x}_{t+1} + \tfrac{\beta_{t}}{M}\|\hvec{x}_{t+1}\|_0 \vec{v}_{t} 
\label{eq:lamp1} ,
\end{eqnarray}
\end{subequations}
with first-layer inputs $\hvec{x}_0=\vec{0}$ and $\vec{v}_0=\vec{y}$.
\Figref{qqplots}(c) shows the QQplot of LAMP's denoiser input error $\big(\hvec{x}_t + \vec{B}_t\vec{v}_t\big)-\vec{x}$. 
The shape of the plot confirms that the error is Gaussian. 

\putFrag{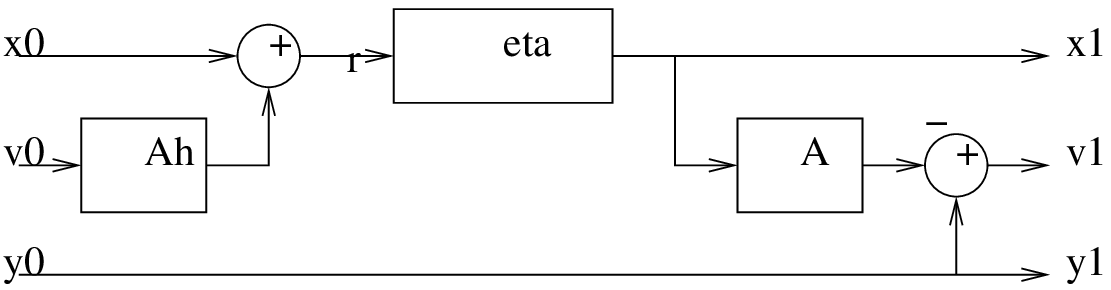}
        {The $t$th layer of the LISTA network, with tunable parameters $\{\vec{A}_t,\vec{B}_t,\lambda_t\}_{t=0}^{T-1}$.}
        {2.3}
        {\newcommand{\sz}{1.0}
         \newcommand{\szz}{0.9}
         \psfrag{+}[c][Bl][\szz]{$+$}
         \psfrag{-}[r][Bl][\szz]{$-$}
         \psfrag{x}[c][Bl][\sz]{$\times$}
         \psfrag{x0}[r][Bl][\sz]{$\hvec{x}_t$}
         \psfrag{x1}[l][Bl][\sz]{$\hvec{x}_{t\!+\!1}$}
         \psfrag{v0}[r][Bl][\sz]{$\vec{v}_t$}
         \psfrag{v1}[l][Bl][\sz]{$\vec{v}_{t\!+\!1}$}
         \psfrag{y0}[r][Bl][\sz]{$\vec{y}$}
         \psfrag{y1}[l][Bl][\sz]{$\vec{y}$}
         \psfrag{r}[t][Bl][\szz]{$\vec{r}_t$}
         \psfrag{Ah}[c][Bl][\sz]{$\vec{B}_t$}
         \psfrag{A}[c][Bl][\sz]{$\vec{A}_t$}
         \psfrag{eta}[c][Bl][\szz]{$\eta(\bullet;\!\lambda_t)$}
        }

\putFrag{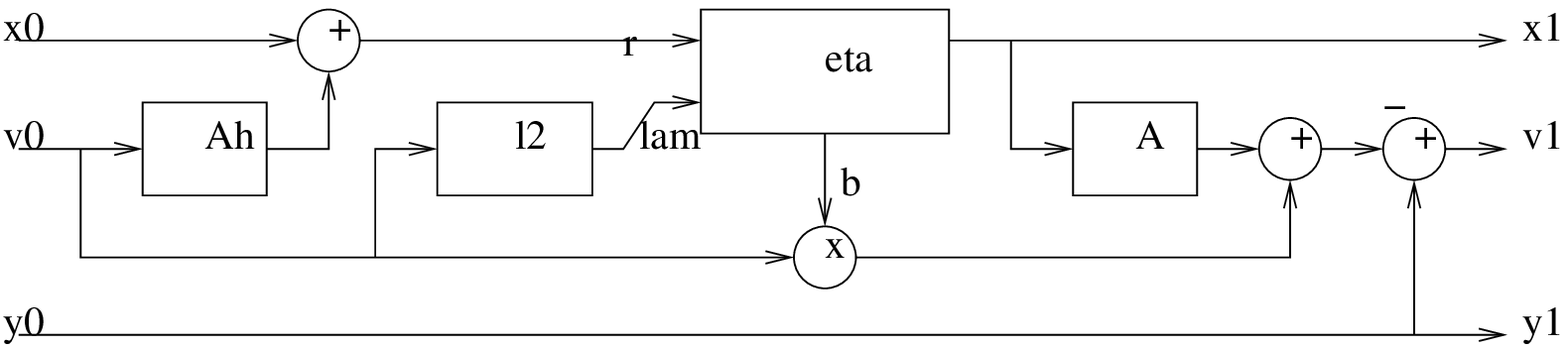}
        {The $t$th layer of the LAMP network, with tunable parameters $\{\vec{A}_t,\vec{B}_t,\alpha_t\}_{t=0}^{T-1}$}
        {3.15}
        {\newcommand{\sz}{1.0}
         \newcommand{\szz}{0.9}
         \newcommand{\szzz}{0.8}
         \psfrag{+}[c][Bl][\sz]{$+$}
         \psfrag{-}[r][Bl][\sz]{$-$}
         \psfrag{x}[c][Bl][\sz]{$\times$}
         \psfrag{x0}[r][Bl][\sz]{$\hvec{x}_t$}
         \psfrag{x1}[l][Bl][\sz]{$\hvec{x}_{t\!+\!1}$}
         \psfrag{v0}[r][Bl][\sz]{$\vec{v}_t$}
         \psfrag{v1}[l][Bl][\sz]{$\vec{v}_{t\!+\!1}$}
         \psfrag{y0}[r][Bl][\sz]{$\vec{y}$}
         \psfrag{y1}[l][Bl][\sz]{$\vec{y}$}
         \psfrag{r}[t][Bl][\szzz]{$\vec{r}_t$}
         \psfrag{l2}[c][Bl][\szzz]{$\frac{\alpha_t\|\bullet\|_2}{\sqrt{M}}$}
         \psfrag{lam}[l][Bl][\szzz]{$\lambda_t$}
         \psfrag{Ah}[c][Bl][\sz]{$\vec{B}_t$}
         \psfrag{A}[c][Bl][\sz]{$\vec{A}_t$}
         \psfrag{eta}[c][Bl][\szz]{$\eta(\bullet;\bullet)$}
         \psfrag{b}[l][Bl][\szzz]{$b_{t+1}$}
        }

\begin{figure}[t]
        \vspace{3mm}
	\newcommand{\wid}{0.3\columnwidth}
	\newcommand{\sz}{0.6}
	\newcommand{\szz}{0.5}
	\begin{tabular}{@{}ccc@{}}
        \psfrag{ISTA}[b][B][\sz]{\sf (a) ISTA}
        \psfrag{Standard Normal Quantiles}[t][t][\szz]{\sf Standard Normal Quantiles}
        \psfrag{Quantiles of Input Sample}[b][b][\szz]{\sf Quantiles of Input Sample}
	\includegraphics[width=\wid]{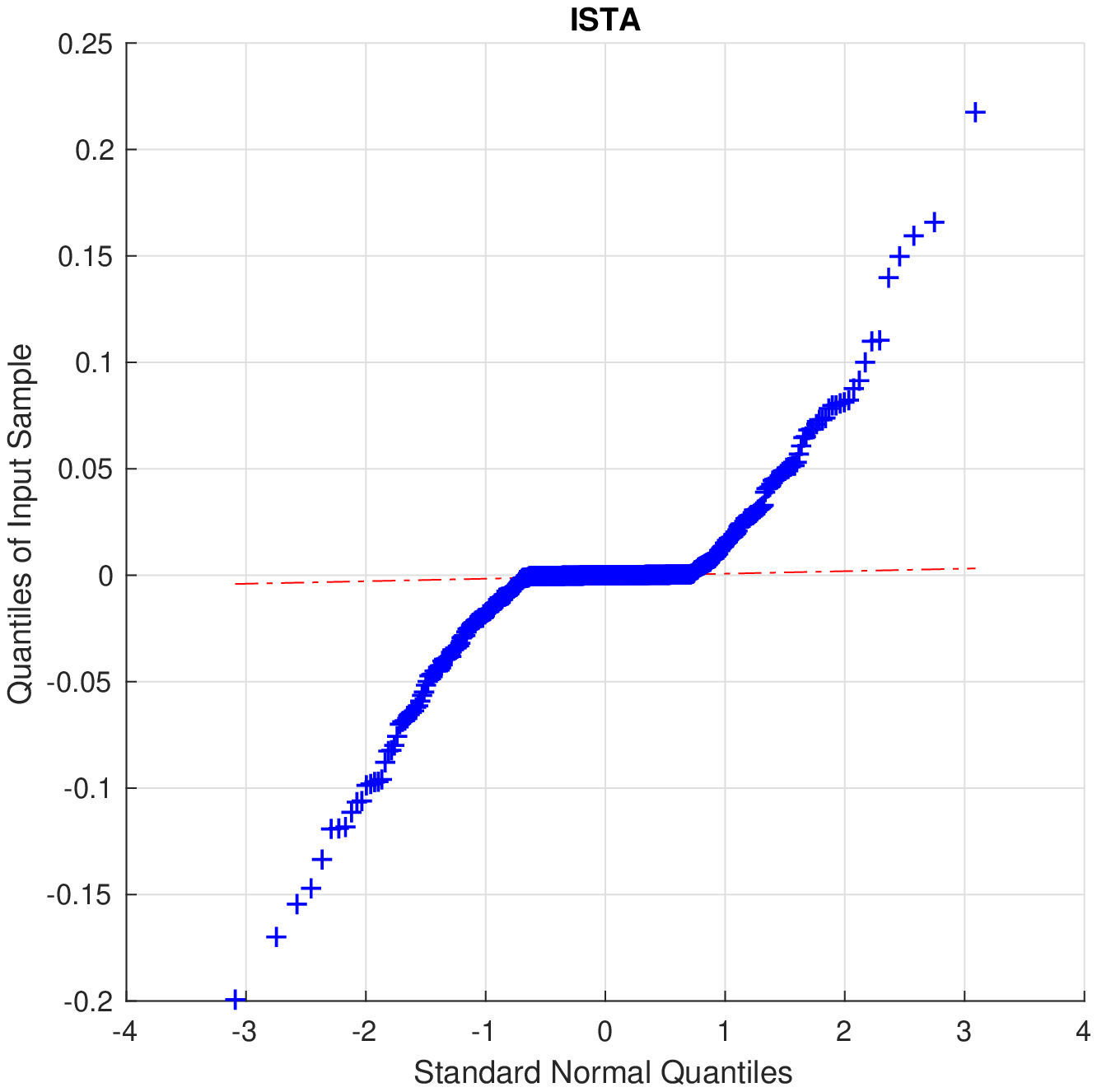}&
        \psfrag{AMP}[b][B][\sz]{\sf (b) AMP}
        \psfrag{Standard Normal Quantiles}[t][t][\szz]{\sf Standard Normal Quantiles}
        \psfrag{Quantiles of Input Sample}[b][b][\szz]{\sf Quantiles of Input Sample}
	\includegraphics[width=\wid]{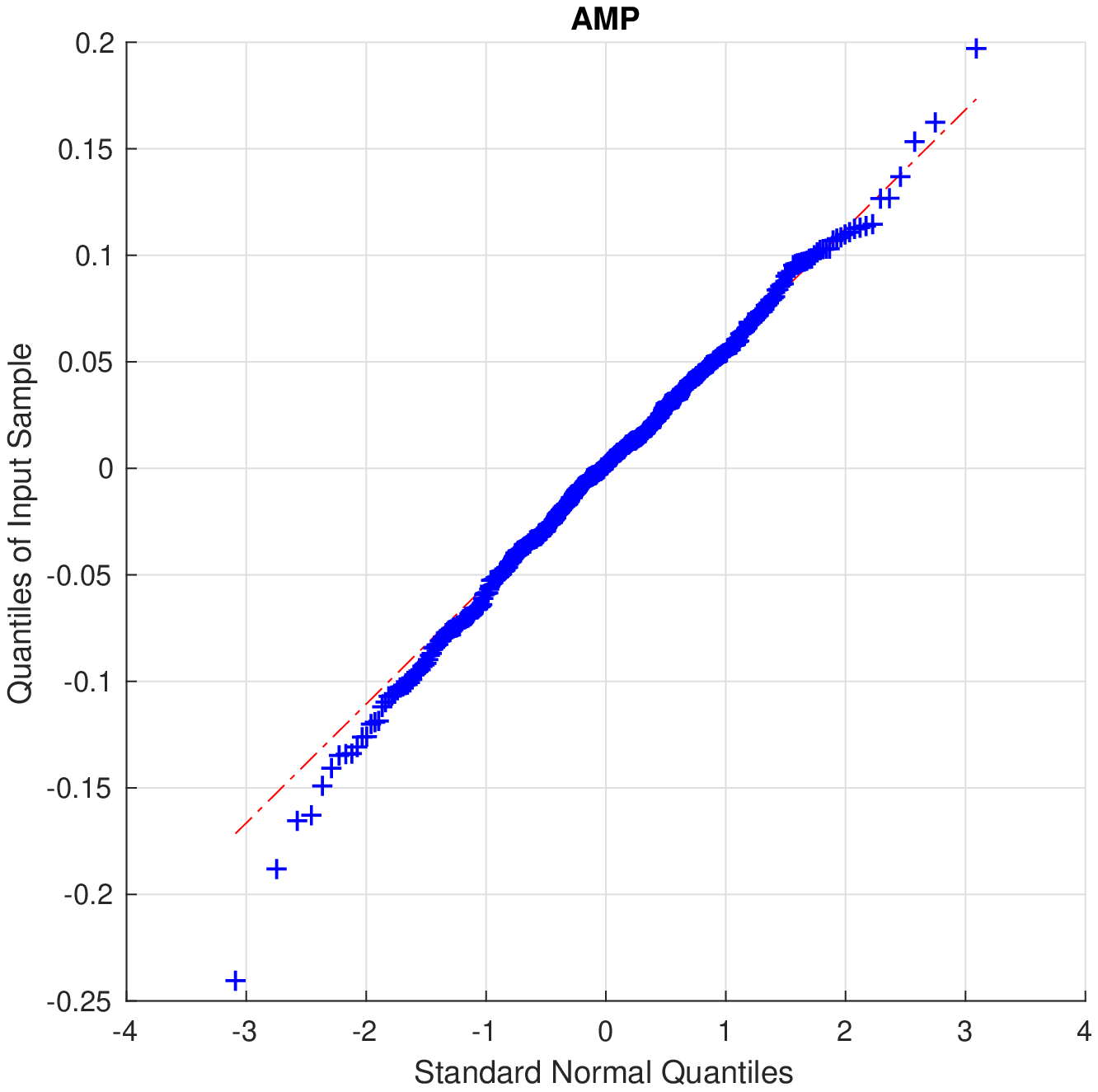}&
        \psfrag{LAMP}[b][B][\sz]{\sf (c) LAMP}
        \psfrag{Standard Normal Quantiles}[t][t][\szz]{\sf Standard Normal Quantiles}
        \psfrag{Quantiles of Input Sample}[b][b][\szz]{\sf Quantiles of Input Sample}
	\includegraphics[width=\wid]{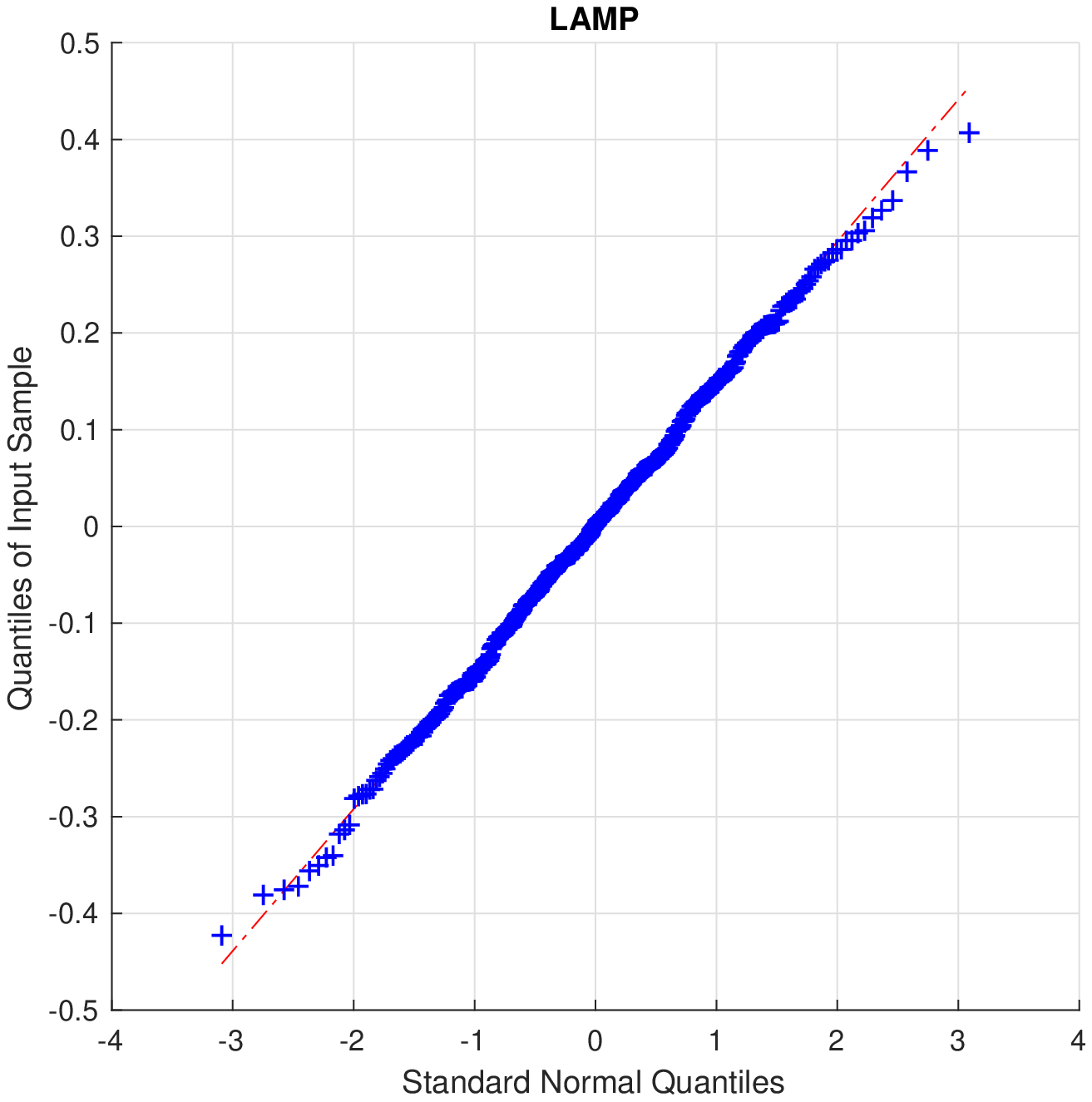}\\
	\end{tabular}
	\caption{QQplots of the denoiser input error evaluated at the first iteration $t$ for which NMSE$(\hvec{x}_t)\!<\!-15$~dB.
        Note ISTA's error is heavy tailed while AMP's and LAMP's errors are Gaussian due to Onsager correction.}
	\label{fig:qqplots}
        \vspace{-3mm}
\end{figure}

\subsection{Learning The LAMP Parameters} \label{sec:learning}

Our proposed learning procedure is described below,
where it is assumed that $\vec{A}$ is known.
But if $\vec{A}$ was unknown, it could be estimated using a least-squares fit to the training data.
Empirically, we find that there is essentially no difference in final test MSE between LAMP networks where i) $\vec{A}$ is known, ii) $\vec{A}$ is estimated via least-squares, or iii) $\vec{A}$ is learned using back-propagation to minimize the loss $\mc{L}_T$ from \eqref{loss}.

\subsubsection{Tied $\{\vec{B}_t\}$} \label{sec:tied}
In the ``tied'' case, $\vec{B}_t=\vec{B}_0~\forall t$, and so the adjustable parameters are $\vec{\Theta}=\big[\vec{B}_0,\{\alpha_t\}_{t=0}^{T-1},\{\beta_t\}_{t=0}^{T-1}\big]$.
We proposed to learn $\vec{\Theta}$ by starting with a $1$-layer network and growing it one layer at a time.
For the first layer (i.e., $t=0$), we
initialize $\vec{B}_0$ as the regularized pseudo-inverse
\begin{align}
\vec{B}_0 = \gamma^{-1}\vec{A}\herm\big(\vec{A}\vec{A}\herm+\vec{I}_M\big)^{-1} 
\label{eq:B0} ,
\end{align}
with $\gamma$ chosen so that $\tr(\vec{A}\vec{B}_0)=N$,
and we use backpropagation to learn the values of $\alpha_0,\beta_0$ that minimize the loss $\mc{L}_0$ from \eqref{loss}.
Then, for each new layer $t=1,\dots,T\!-\!1$, we 
\begin{enumerate}
\item initialize $\vec{B}_0,\alpha_t,\beta_t$ at the values from layer $t\!-\!1$, 
\item optimize $\alpha_t,\beta_t$ alone using backpropagation, and 
\item re-optimize all parameters $\vec{B}_0,\{\alpha_i\}_{i=0}^t,\{\beta_i\}_{i=0}^t$ using backpropagation to minimize the loss $\mc{L}_t$ from \eqref{loss}.
\end{enumerate}

\subsubsection{Untied $\{\vec{B}_t\}$} \label{sec:untied}
In the ``untied'' case, $\vec{B}_t$ is allowed to vary across layers $t$.
We propose the same learning procedure as above, but with one exception: 
when initializing the parameters for each new layer, we initialize $\vec{B}_t$ at the regularized pseudo-inverse \eqref{B0} rather than the learned value $\vec{B}_{t-1}$. 

\subsubsection{Structured $\vec{B}_t$}
Motivated by the $M\!<\!N$ case, we recommend constraining $\vec{B}_t=\vec{A}\herm\vec{C}_t$ and learning only $\vec{C}_t\in\Complex^{M\times M}$, which reduces the number of free parameters in $\vec{B}_t$ from $MN$ to $M^2$.
After these $\vec{B}_t$ are learned, there is no need to represent them in factored form, although doing so may be advantageous if $\vec{A}$ has a fast implementation (e.g., FFT). 
Empirically, we find that there is essentially no difference in final test MSE between LAMP networks where i) $\vec{B}_t$ is unconstrained and ii) $\vec{B}_t=\vec{A}\herm\vec{C}_t$.
However, the learning procedure is more efficient in the latter case.

\section{Numerical Results} \label{sec:numerical}

We now evaluate the performance of LISTA and LAMP on the sparse linear inverse problem described in \secref{compare}.
The data were generated as described in \secref{compare}, with
training mini-batches of size $D\!=\!1000$ and a single testing mini-batch of size $1000$ (drawn independently of the training data).
The training and testing methods were implemented in Python using TensorFlow \cite{Abadi:tensorflow:15} with the ADAM optimizer \cite{Kingma:ICLR:15}.
For LAMP, we performed the learning as described in \secref{learning}.
For LISTA, we used the same approach to learn ``untied''
$\big[\vec{B}_0,\vec{S}_0,\{\lambda_t\}_{t=0}^{T-1}\big]$
and ``tied'' $\{\vec{B}_t,\vec{S}_t,\lambda_t\}_{t=0}^{T-1}$,
with no constraints on $\vec{S}_t$ or $\vec{B}_t$ (because we found that adding constraints degraded performance).

\Figref{lista_vs_lamp_Giid} shows average test NMSE $\|\hvec{x}_t-\vec{x}\|_2^2/\|\vec{x}\|^2$ versus layer $t$ for the same i.i.d.\ Gaussian $\vec{A}$ and test data used to create \figref{ista_fista_amp_phil}, allowing a direct comparison.
The figure shows LAMP significantly outperforming LISTA and AMP in convergence time and final NMSE.
For example, to reach $-34$~dB NMSE,
tied-LAMP took only $7$ layers, 
tied-LISTA took $15$, 
and AMP took 23.

\Figref{lista_vs_lamp_k15} shows the results of a similar experiment, but where the singular values of $\vec{A}$ were replaced by a geometric sequence that yielded $\|\vec{A}\|_F^2=N$ and a condition-number of $15$.
For this $\vec{A}$, AMP diverged but LAMP did not, due in part to the ``preconditioning'' effect of \eqref{B0}.

The figures also show that the untied versions of LAMP and LISTA yielded small improvements over the tied versions, 
but at the cost of a $T$-fold increase in parameter storage plus significantly increased training time.

\putFrag{lista_vs_lamp_Giid}
	{Test NMSE versus layer for i.i.d.\ Gaussian $\vec{A}$.}
	{\figsize}
	{\newcommand{\sz}{0.5}
	 \newcommand{\szz}{0.7}
         \psfrag{LISTA}[l][l][\sz]{\sf LISTA tied}
         \psfrag{LISTA untied}[l][l][\sz]{\sf LISTA untied}
         \psfrag{LAMP}[l][l][\sz]{\sf LAMP tied}
         \psfrag{LAMP untied}[l][l][\sz]{\sf LAMP untied}
         \psfrag{NMSE (dB)}[][][\szz]{\sf average NMSE [dB]}
         \psfrag{Layers}[][][\szz]{\sf layers}
         }

\putFrag{lista_vs_lamp_k15}
	{Test NMSE versus layer for $\vec{A}$ with condition number $15$.}
	{\figsize}
	{\newcommand{\sz}{0.51}
	 \newcommand{\szz}{0.7}
         \psfrag{LISTA tied}[l][l][\sz]{\sf LISTA tied}
         \psfrag{LISTA untied}[l][l][\sz]{\sf LISTA untied}
         \psfrag{LAMP}[l][l][\sz]{\sf LAMP tied}
         \psfrag{LAMP untied}[l][l][\sz]{\sf LAMP untied}
         \psfrag{NMSE (dB)}[][][\szz]{\sf average NMSE [dB]}
         \psfrag{Layers}[][][\szz]{\sf layers}
         }

\section{Conclusion}
We considered the application of deep learning to the sparse linear inverse problem. 
Motivated by the AMP algorithm, we proposed the use of Onsager correction in deep neural networks, for the purpose of decoupling and Gaussianizing errors across layers. 
Empirical results demonstrated improved accuracy and efficiency over Gregor and LeCun's LISTA \cite{Gregor:ICML:10}.

\clearpage
\bibliographystyle{ieeetr}
\bibliography{macros_abbrev,books,misc,comm,multicarrier,sparse,machine}

\begin{thebibliography}{10}

\bibitem{Eldar:Book:12}
Y.~C. Eldar and G.~Kutyniok, {\em Compressed Sensing: Theory and Applications}.
\newblock New York: Cambridge Univ. Press, 2012.

\bibitem{Olshausen:VR:97}
B.~A. Olshausen and D.~J. Field, ``Sparse coding with an overcomplete basis
  set: {A} strategy employed by v1,'' {\em Vision Research}, vol.~37,
  pp.~3311--3325, 1997.

\bibitem{Gregor:ICML:10}
K.~Gregor and Y.~{LeCun}, ``Learning fast approximations of sparse coding,'' in
  {\em Proc. Int. Conf. Mach. Learning}, pp.~399--406, 2010.

\bibitem{Sprechmann:ICML:12}
P.~Sprechmann, P.~Bronstein, and G.~Sapiro, ``Learning efficient
  structured-sparse models,'' in {\em Proc. Int. Conf. Mach. Learning},
  pp.~615--622, 2012.

\bibitem{Kamilov:SPL:16}
U.~Kamilov and H.~Mansour, ``Learning optimal nonlinearities for iterative
  thresholding algorithms,'' {\em IEEE Signal Process. Lett.}, vol.~23,
  pp.~747--751, May 2016.

\bibitem{Wang:AAAI:16}
Z.~Wang, Q.~Ling, and T.~S. Huang, ``Learning deep $\ell_0$ encoders,'' in {\em
  Proc. AAAI Conf. Artificial Intell.}, pp.~2194--2200, 2016.

\bibitem{Donoho:PNAS:09}
D.~L. Donoho, A.~Maleki, and A.~Montanari, ``Message passing algorithms for
  compressed sensing,'' {\em Proc. Nat. Acad. Sci.}, vol.~106,
  pp.~18914--18919, Nov. 2009.

\bibitem{Tibshirani:JRSSb:96}
R.~Tibshirani, ``Regression shrinkage and selection via the lasso,'' {\em J.
  Roy. Statist. Soc. B}, vol.~58, no.~1, pp.~267--288, 1996.

\bibitem{Chen:JSC:98}
S.~S. Chen, D.~L. Donoho, and M.~A. Saunders, ``Atomic decomposition by basis
  pursuit,'' {\em SIAM J. Scientific Comput.}, vol.~20, no.~1, pp.~33--61,
  1998.

\bibitem{Candes:CPAM:06}
E.~Cand\`{e}s, J.~Romberg, and T.~Tao, ``Stable signal recovery from incomplete
  and inaccurate measurements,'' {\em Communications on Pure and Applied
  Mathematics}, vol.~59, no.~8, pp.~1207--1223, 2006.

\bibitem{Chambolle:TIP:98}
A.~Chambolle, R.~A. DeVore, N.~Lee, and B.~J. Lucier, ``Nonlinear wavelet image
  processing: {V}ariational problems, compression, and noise removal through
  wavelet shrinkage,'' {\em IEEE Trans. Image Process.}, vol.~7, pp.~319--335,
  Mar. 1998.

\bibitem{Daubechies:CPAM:04}
I.~Daubechies, M.~Defrise, and C.~D. Mol, ``An iterative thresholding algorithm
  for linear inverse problems with a sparsity constraint,,'' {\em Commun. Pure
  \& Appl. Math.}, vol.~57, pp.~1413--1457, Nov. 2004.

\bibitem{Beck:JIS:09}
A.~Beck and M.~Teboulle, ``A fast iterative shrinkage-thresholding algorithm
  for linear inverse problems,'' {\em SIAM J. Imag. Sci.}, vol.~2, no.~1,
  pp.~183--202, 2009.

\bibitem{Montanari:Chap:12}
A.~Montanari, ``Graphical models concepts in compressed sensing,'' in {\em
  Compressed Sensing: Theory and Applications} (Y.~C. Eldar and G.~Kutyniok,
  eds.), Cambridge Univ. Press, 2012.

\bibitem{Bayati:TIT:11}
M.~Bayati and A.~Montanari, ``The dynamics of message passing on dense graphs,
  with applications to compressed sensing,'' {\em IEEE Trans. Inform. Theory},
  vol.~57, pp.~764--785, Feb. 2011.

\bibitem{Donoho:ITW:10a}
D.~L. Donoho, A.~Maleki, and A.~Montanari, ``Message passing algorithms for
  compressed sensing: {I. M}otivation and construction,'' in {\em Proc. Inform.
  Theory Workshop}, (Cairo, Egypt), pp.~1--5, Jan. 2010.

\bibitem{Goodfellow:Book:16}
I.~Goodfellow, Y.~Bengio, and A.~Courville, {\em Deep Learning}.
\newblock MIT Press, 2016.

\bibitem{Hershey:Tech:14}
J.~R. Hershey, J.~{Le Roux}, and F.~Weninger, ``Deep unfolding: {M}odel-based
  inspiration of novel deep architectures,'' Tech. Rep. TR2014-117, Mitsubishi
  Electric Research Labs, 2014.
\newblock (See also arXiv:1409.2574).

\bibitem{Schmidt:CVPR:14}
U.~Schmidt and S.~Roth, ``Shrinkage fields for effective image restoration,''
  in {\em Proc. IEEE Conf. Comp. Vision Pattern Recog.}, pp.~2774--2781, 2014.

\bibitem{Dong:TPAMI:16}
C.~Dong, C.~C. Loy, K.~He, and X.~Tang, ``Image super-resolution using deep
  convolutional networks,'' {\em IEEE Trans. Pattern Anal. Mach. Intell.},
  vol.~38, pp.~295--307, Feb. 2016.

\bibitem{Mousavi:ALL:15}
A.~Mousavi, A.~Patel, and R.~Baraniuk, ``A deep learning approach to structured
  signal recovery,'' in {\em Proc. Allerton Conf. Commun. Control Comput.},
  pp.~1336--1343, 2015.

\bibitem{Kulkarni:CVPR:16}
K.~Kulkarni, S.~Lohi, P.~Turaga, R.~Kerviche, and A.~Ashok, ``{ReconNet:
  N}on-iterative reconstruction of images from compressively sensed random
  measurements,'' in {\em Proc. IEEE Conf. Comp. Vision Pattern Recog.}, 2016.
\newblock (see also arXiv:1601.06892).

\bibitem{Iliadis:16}
M.~Iliadis, L.~Spinoulas, and A.~K. Katsaggelos, ``Deep fully-connected
  networks for video compressive sensing,'' in {\em arXiv:1603:04930}, 2016.

\bibitem{Abadi:tensorflow:15}
M.~Abadi, A.~Agarwal, P.~Barham, {\em et~al.}, ``{TensorFlow: L}arge-scale
  machine learning on heterogeneous systems,'' 2015.
\newblock Software available from tensorflow.org.

\bibitem{Kingma:ICLR:15}
D.~P. Kingma and J.~Ba, ``Adam: {A} method for stochastic optimization,'' in
  {\em Proc. Internat. Conf. on Learning Repres.}, 2015.
\newblock (see also arXiv:1412.6980).

\end{thebibliography}

\end{document}